\documentclass[epj]{svjour}
\usepackage{graphicx}
\usepackage{bm}   
\usepackage{color}
\usepackage{amssymb}
\usepackage{amsmath}
\usepackage{balance}
\usepackage{subfigure}
\usepackage{enumerate}
\usepackage{widetext}

\begin{document}
	
	\title{High-order synchronization in a system of nonlinearly coupled Stuart-Landau oscillators}
	
	\author{Nissi Thomas \inst{1}, S.~Karthiga \inst{1} \and M. Senthilvelan \inst{1,} \thanks{\emph{e-mail:velan@cnld.bdu.ac.in}}}
	\institute{Department of Nonlinear Dynamics, School of Physics, Bharathidasan University, Tiruchirappalli 620024, Tamil Nadu, India.}

	\abstract{
	The high-order synchronization was studied in systems driven by external force and in autonomous systems with proper frequency mismatch. Differing from the literature, in this article, we demonstrate the occurrence of high-order ($1:2$) synchronization in an autonomous nonlinearly coupled (Stuart-Landau) oscillators which admit a particular form of rotational symmetry. Interestingly, the observed $1:2$ synchronization happens not only for a particular choice of natural frequencies but for all possible choices of frequencies.   We have observed such a behaviour in the case of $1:1$ synchronization, where we have seen a variety of couplings in the literature that forces the oscillators to have almost equal frequencies and makes the system to oscillate with a common frequency independent of whether the oscillators are identical or non-identical.  Similarly, in this article we observe that, whether the initial choice of frequencies is of the ratio $1:2$ or not, the given nonlinear coupling forces the system to oscillate with the frequency ratio $1:2$.  Further, we present synchronization and other dynamical behaviours of the system by considereing different choices of natural frequencies.}
	\titlerunning{High-order synchronization in a system of nonlinearly coupled Stuart-Landau oscillators}
	\authorrunning{Nissi Thomas et al.}
	\maketitle

\section{Introduction}
\par Synchronization, one of the fascinating phenomena observed in coupled systems where this subject has seen a very wide literature \cite{pikovsky}. Various types of entrainment behaviour \cite{ha,um,korono,park} and various forms of synchrony patterns \cite{krishna,kuramo,panag,preml,mais,rybalo,kang}  have been studied in different systems with different coupling topologies.  
When two or more oscillators are coupled, they may start to oscillate with the same frequency and their amplitudes and phases may also get related. Such oscillations with equalized frequency with suitable coupling is known as $1:1$ synchronization  \cite{pikovsky} and there are variety of physical systems which exhibit $1:1$ synchronization.  Besides this, high-order synchronization, in which the frequencies of the oscillators are related by a particular ratio ($n:m$), may also occur \cite{pikovsky}.   Comparing the literatures of $1:1$ and $n:m$ synchronization, the latter is less studied but still the occurence of such synchronization is observed in physical and biological systems including   cardiorespiratory system \cite{schafer},  the josephson junction based electrical rotator \cite{jain}, a ruby nuclear magnetic resonance laser with delayed feedback \cite{simonet} and in the circuit of thermally coupled VO$_2$ (vanadium dioxide) oscillators \cite{velichi1}.  Besides this, the possibility of using high-order synchronized states in reservior computing is also  demonstrated recently in the Refs.  \cite{velichi1,velichi2,velichi3,velichi4,velichi5}.
\par  It is evident from the literature that the high-order synchronization has been explored much in the case of forced or non-autonomous systems.  In the case of autonomous systems, this $n:m$ synchronization is achieved through frequency mismatch only when the natural frequency of the oscillators is said to have the ratio $\omega_1:\omega_2$$=n:m$ \cite{pikovsky}.  In this article we report the occurrence of $n:m$ synchronization in an autonomous system of coupled Stuart-Landau oscillators where the nonlinear coupling among the oscillators is shown to induce $1:2$ synchronization.   A symmetry admitted by this nonlinear coupled system is responsible for the occurence of this high-order ($1:2$) synchronization.  Because of this symmetry, there is no restriction imposed on the natural frequencies of the oscillators.  In other words, whatever be the natural frequencies of the two oscillators (whether they are identical or non-identical or their natural frequencies have the ratio $1:2$) the symmetry admitted by the system only allows $1:2$ synchronization and no other type of synchronization. By considering the dynamical behaviours of the system for different cases of $\omega_1$ and $\omega_2$, we illustrate both analytically and numerically that all the stationary states of the system represent $1:2$ synchronized state alone and no stationary states representing any other synchronized state do exist.  We also bring out few other interesting dynamical aspects of the system under this nonlinear coupling. The first and foremost is that the system shows a multistable behaviour when the coupling strength is varied. Due to this, the system exhibits oscillations in both clockwise and anticlockwise direction.   We also present an interesting case in which the multistable clockwise and anticlockwise states have the same periodic orbit with same amplitude but different frequencies of oscillation.
\par We organize our work as follows. In Sec. \ref{sec2}, we present the considered model and discuss briefly the possibility of practical realization of the model.  In Sec. \ref{sec3}, we discuss the dynamics for the case $\omega_2 \neq 2\omega_1$ where we elucidate the results with two different cases, namely (i) $\omega_1=\omega_2$ and (ii) $\omega_2 \neq \omega_1$.  We also analyze the stationary states corresponding to these two cases and discuss their  stabilities in this section. In Sec. \ref{sec4}, we investigate the dynamics of the system when the natural frequencies of the uncoupled oscillators have the frequency relation $\omega_1:\omega_2=1:2$.  Finally, we summarize our results in Sec. \ref{sec5}.

\section{\label{sec2}Model}
We consider two Stuart-Landau oscillators interacting with each other through an asymmetric quadratic coupling in the following manner,
\begin{eqnarray}
\dot \alpha =&(\gamma_1-\gamma_2|\alpha|^2)\alpha -i \omega_1 \alpha-2i\zeta\alpha^*\beta,  \nonumber \\
\dot \beta=&(\gamma_1-\gamma_2|\beta|^2)\beta-i \omega_2 \beta-i\zeta\alpha^2, \label{beta}
\end{eqnarray}
where $\alpha=x_1+iy_1 \in \mathbb{C}$ and $\beta=x_2+iy_2 \in \mathbb{C}$ are the state variables corresponding to the two Stuart-Landau oscillators. $\alpha^*$ and $\beta^*$ represent their complex conjugates.  The parameters $\gamma_1 \in {\cal{R}}^+$ and $\gamma_2 \in {\cal R}^+$ respectively represent the negative linear damping (linear gain) and positive nonlinear damping (nonlinear loss) strengths. The damping rate $\gamma_1$ is also known as Hopf bifurcation parameter as the system executes steady state dynamics for $\gamma_1<0$, supercritical Hopf bifurcation at $\gamma_1=0$ and oscillatory dynamics for $\gamma_1>0$.  The parameters  $\omega_1$ and $\omega_2$ respectively denote natural frequencies of the two uncoupled Stuart-Landau oscillators and $\zeta$ represents the strength of the asymmetric nonlinear coupling.  As the oscillators $\alpha$ and $\beta$ are coupled to each other in an assimilar way (vide Eq. (\ref{beta})), we call the coupling as asymmetric coupling.  
\par This type of coupling may be achieved in different contexts.  In a very recent work \cite{nissi}, two of the present authors have shown the possibility of observing the above system (\ref{beta}) (or the nonlinear coupling) in the context of trapped ions.  In this recent article, the Lindblad master equation corresponding to a system of two quantum van der Pol oscillators coupled via a nonlinear cross-Kerr interaction has been considered. This nonlinear interaction can be achieved by considering the strong coulombic interaction between two trapped-ions \cite{ding}. The classical amplitude equations obtained from the Lindbland master equation of the coupled system is similar to the one given in Eq.(\ref{beta}).
\section{\label{sec3}Dynamics of the system\\ (Case: $\omega_2 \neq 2 \omega_1$)}{\label{III}}
\subsection{Stationary periodic orbits}{\label{III_a}}

\par Now we unveil the dynamical features that come out due to the  nonlinear coupling.  We recall here that in the absence of coupling, the two Stuart-Landau oscillators execute oscillatory dynamics along clockwise direction with amplitude $\sqrt{\gamma_1/\gamma_2}$ with respective frequencies $\omega_1$ and $\omega_2$. In the presence of coupling, we can study the dynamics of the system qualitatively by examining the stability of their stationary states or modes.   For this purpose,  by considering $\alpha=R_{1}\exp{(i\theta_1)}$ and $\beta=R_{2}\exp{(i\theta_2)}$, we can obtain the dynamical equations for amplitudes $R_i$ and phases $\theta_i$ ($i=1$ and $2$) corresponding to Eq. (\ref{beta}) and they are given by 
\begin{eqnarray}
\dot{R}_1&=&(\gamma_1-\gamma_2 R_1^2)R_1+2\zeta R_1R_2 \sin(\theta_2 -2 \theta_1),  \nonumber \\
\dot{R}_2&=&(\gamma_1-\gamma_2 R_2^2)R_2-\zeta R_1^2 \sin(\theta_2 -2 \theta_1), \nonumber\\
\dot{\theta}_1&=&-\omega_1 - 2 \zeta R_2 \cos(\theta_2 -2 \theta_1), \nonumber \\
\dot{\theta}_2&=&-\omega_2- \zeta \frac{R_1^2}{R_2} \cos(\theta_2-2 \theta_1). \label{amp_phase}
\end{eqnarray}
 \par A  closer observation on Eq. (\ref{amp_phase}) reveals that the system remains same with repect to the transformation $\theta_2 \rightarrow \theta_2+2 \psi$ and $\theta_1 \rightarrow \theta_1+\psi$.  Further discussion will make clear that this symmetric nature is responsible for the $1:2$ type synchronization in the system (in the original Eq. (\ref{beta}), this symmetry can be seen as the invariance of the system with respect to the transformation $\alpha \rightarrow \alpha e^{i \psi}$ and $\beta \rightarrow \beta e^{2 i \psi}$).  We can reduce the above Eq. (\ref{amp_phase}) to 
\begin{eqnarray}
\dot{R}_1&=&(\gamma_1-\gamma_2R_1^2)R_1+2\zeta R_1R_2 \sin \delta,  \nonumber \label{r11}\\
\dot{R}_2&=&(\gamma_1-\gamma_2R_2^2)R_2-\zeta R_1^2\sin\delta, \label{r21} \nonumber \\
\dot{\delta}&=&-\Delta-\zeta\big(\frac{R_1^2-4R_2^2}{R_2}\big)\cos\delta, \label{delta1}
\end{eqnarray}
where $\delta=\theta_2-2\theta_1$ is a form of phase difference and $\Delta=\omega_2-2\omega_1$ represents the frequency detuning among the two Stuart-Landau oscillators.  The stationary oscillatory states or limit cycles of the system can be found by setting $\dot R_1=\dot R_2=\dot \delta=0$.  In the following, we present the form of the stationary periodic orbits corresponding to all the cases excluding $\omega_2 = 2 \omega_1$. The dynamics and the possible stable and unstable periodic orbits or limit cycles corresponding to the limiting case $\omega_2 = 2 \omega_1$ are discussed separately in Sec. \ref{sec4}.  For the case $\omega_2 \neq 2 \omega_1$, the system supports five possible periodic states which are given by
\begin{eqnarray}
R_{1}^{*}=\sqrt{\chi}R_{2}^{*},  \quad R_{2}^{*}=\sqrt{\frac{\gamma_1}{\gamma_2}\frac{(\chi+2)}{(\chi^{2}+2)}}, \nonumber \\ \delta^{*}=\tan^{-1}\left[-\frac{\gamma_2R_2^{*2}}{\Delta}\frac{(\chi-1)(\chi-4)}{(\chi+2)}\right],\label{ss1}
\end{eqnarray}
where $\chi$ can be determined by solving the quintic polynomial equation
{\small\begin{equation}
\frac{\gamma_1}{\gamma_2}\zeta^{2}(\chi+2)(\chi^{2}+2)(\chi-4)^{2}-\gamma_1^{2}(\chi-1)^{2}(\chi-4)^{2}-\Delta^{2}(\chi^{2}+2)^{2}=0\label{fpp}.
\end{equation}}
\par The above quintic polynomial has five roots in which the real and positive roots indicate stationary periodic orbits.  Due to the non-trivial forms of the roots, their explicit expressions are not presented here. The stability nature of the obtained periodic orbits can be determined from the Jacobian matrix
\begin{widetext}
	\begin{equation}
	J=\begin{bmatrix} 
	\gamma_1-3\gamma_1R_1^{*2}+2\zeta R_2^*\sin\delta^* & 2\zeta R_1^*\sin{\delta^*} & 2\zeta R_1^*R_2^*\cos\delta^* \\
	-2\zeta R_1^*\sin\delta^*& \gamma_1-3\gamma_2R_2^{*2} & -\zeta R_1^{*2}\cos\delta^*\\
	-2\zeta\frac{R_1^*}{R_2^*} \cos\delta^*& \zeta\left(\frac{R_1^{*2}+4R_2^{*2}}{R_2^{*2}}\right)\cos\delta^* & \zeta\left(\frac{R_1^{*2}-4R_2^{*2}}{R_2^{*}}\right)\sin\delta^* \\ 
	\end{bmatrix}.
	\end{equation}{\label{jacob}}
\end{widetext}

We have analyzed all five cases. Our results indicate that there exists two possible periodic orbits which become stable in different parametric regions. Before we discuss the results of the linear stability analysis corresponding to different cases, we wish to indicate certain important inferences that we can observe from the form of the above stationary states. 
\par The first observation one can make from Eq. (\ref{ss1}) is that in none of the stationary states (mentioned in Eq. (\ref{ss1}) and (\ref{fpp})), the amplitude of the two oscillators match with each other ($R_1^* \neq R_2^*$) and their ratio is found to be equal to $\sqrt{\chi}$ (where $\chi$ takes the value $1$ only at a particular parametric value and not for a range of parametric values (or for a range of values of $\zeta$)).  Similarly, considering the frequency of the oscillators $\alpha$ and $\beta$ in these stationary states (the underlying expressions can be deduced by substituting the values of $R_1^*$, $R_2^*$, and $\delta^*$ corresponding to the stationary states (Eq. (\ref{ss1}) and (\ref{fpp})) in the expressions of $\dot{\theta_1}$ and $\dot{\theta_2}$ of Eq. (\ref{amp_phase})), we observe that they take the values,
\begin{eqnarray}
\dot{\theta}_1^*&=-\omega_1-2\zeta R_2^*\cos\delta^*=-\omega_1+2\Delta \displaystyle{\frac{1}{\chi-4}}, \nonumber \\
\dot{\theta}_2^*&=-\omega_2-\zeta\displaystyle{\frac{R_1^{*2}}{R_2^*}}\cos\delta^*=-\omega_2+ \Delta \displaystyle{\frac{\chi}{\chi-4}}. \label{tstar}
\end{eqnarray}
Equation (\ref{tstar}) clearly indicates that
\begin{equation}
\dot{\theta}_1^* \neq \dot{\theta}_2^*,
\end{equation}
and so none of the states mentioned in Eq. (\ref{ss1}) show amplitude as well as frequency matching among the two oscillators ($\alpha$ and $\beta$).  It is a well known fact that at the asymptotic time limit, the dynamical system approaches (or settles into) one of its stable stationary configurations.  In this regard, the above discussion says that none of the stationary states (including unstable and stable states) correspond to first order or $1:1$ synchronized state.  This mimics that the introduced nonlinear coupling (\ref{beta}) may not induce first order synchronization in the considered system.  A closer look on Eq. (\ref{tstar}) (and use $\Delta =\omega_2-2 \omega_1$) further reveals the role of coupling where all the stationary states satisfy
\begin{equation}
\displaystyle{\frac{\dot{\theta}_2^*}{\dot{\theta}_1^*}}=2.   \label{rat1}
\end{equation}
Thus, the stationary states of the system represent the state of $1:2$ synchronization where the angular frequency of the second oscillator is twice the angular frequency of the first oscillator.  Hence the introduced coupling facilitates higher-order (or $1:2$) synchronization rather than first order synchronization.  Note that the relation (\ref{rat1}) holds for all choices of $\omega_1$ and $\omega_2$ and so only $1:2$ synchronization is observed in all the cases.  Thus, the relation (\ref{rat1}) clearly indicates the usefulness of the considered asymmetric coupling in inducing high order synchronization.    To illustrate it more clearly, we discuss two important cases in the following sub-sections.  
\subsection{Case (i): $\omega_1=\omega_2$}
\par
\begin{figure*}[htb!]
	\centering
	\includegraphics[width=0.85\linewidth]{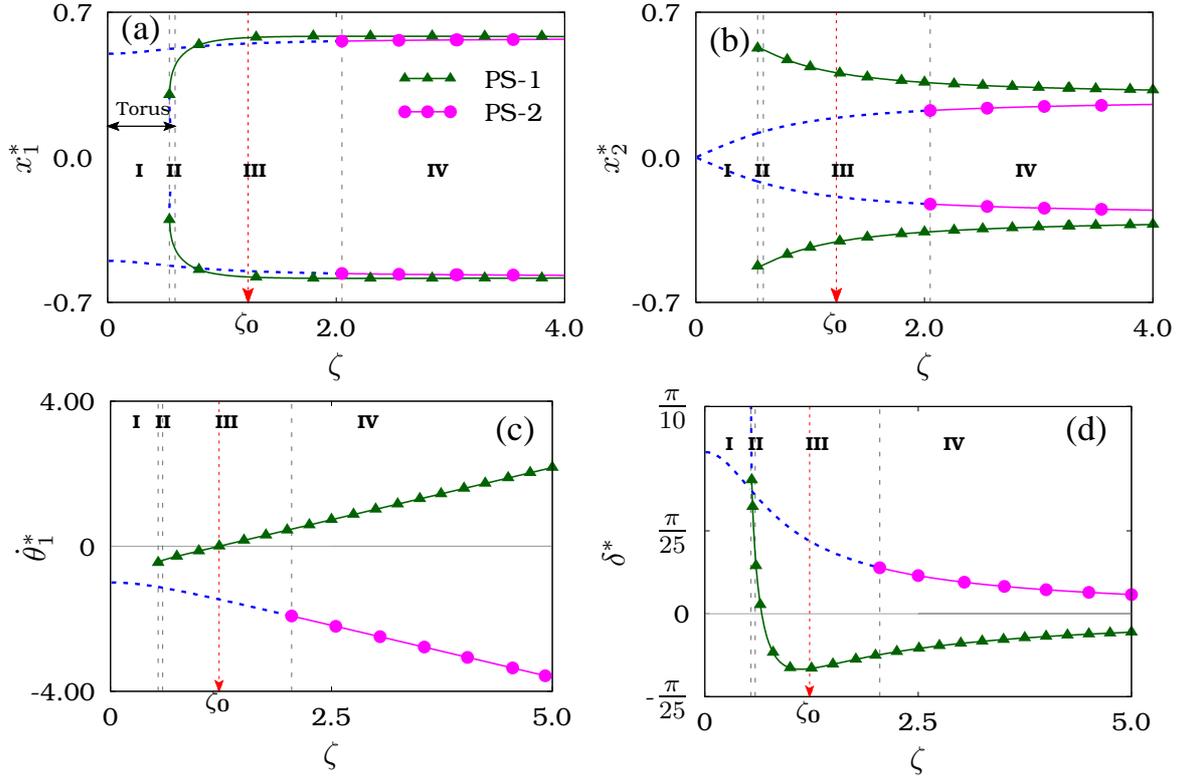}
	\caption{Maxima and minima values of the variables $x_1$ and $x_2$ in different stable and unstable stationary states of the system in Eq. (\ref{beta}) for $\omega_1=\omega_2=1.0$ are shown respectively in Figs. \ref{fig1}(a) and \ref{fig1}(b) for different values of $\zeta$.  Similarly, the angular frequency $\dot{\theta_1}^*$ and the relative phase difference $\delta^*$ corresponding to different stable and unstable states are shown with respect to $\zeta$ in Figs. \ref{fig1}(c) and \ref{fig1}(d).  In all the above figures, we represent the unstable states with dashed lines and the stable states with continuous line with some filled  attributes like filled circles or filled triangles. In all these figures, we consider $\gamma_1=0.25$, $\gamma_2=1.0$ and $\omega_1=\omega_2=1.0$. }
	\label{fig1}
\end{figure*}

 As a first case, we consider a simple situation in which two identical oscillators (that is, $\omega_1=\omega_2=\omega=1.0$) are coupled. In the absence of coupling, due to the identical nature, the two oscillators oscillate with the same frequency and amplitude.  However, this is not possible in the presence of asymmetric coupling and the system will no longer support stationary periodic orbits with identical frequency and amplitude.   The dynamical characteristics of this case is illustrated in Fig. \ref{fig1} where the information of the amplitude, frequency and relative phase of the two oscillators are given with respect to $\zeta$.  The maxima and minima values of $x_1$ and $x_2$ (corresponding to different stable or unstable stationary periodic orbits) are obtained using Eqs. (\ref{ss1}) and (\ref{fpp}) and are presented in Figs. \ref{fig1}(a) and \ref{fig1}(b).  Similarly, the angular frequency of the first oscillator, $\dot{\theta}_1^*$, (the frequency of the second oscillator, $\dot{\theta}_2^*=2 \dot{\theta}_1^*$) and the relative phase difference $\delta^*$ corresponding to different stable and unstable orbits are shown respectively in Figs. \ref{fig1}(c) and \ref{fig1}(d). 
\par
\begin{figure*}[h]
	\begin{center}
		\includegraphics[width=0.95\linewidth]{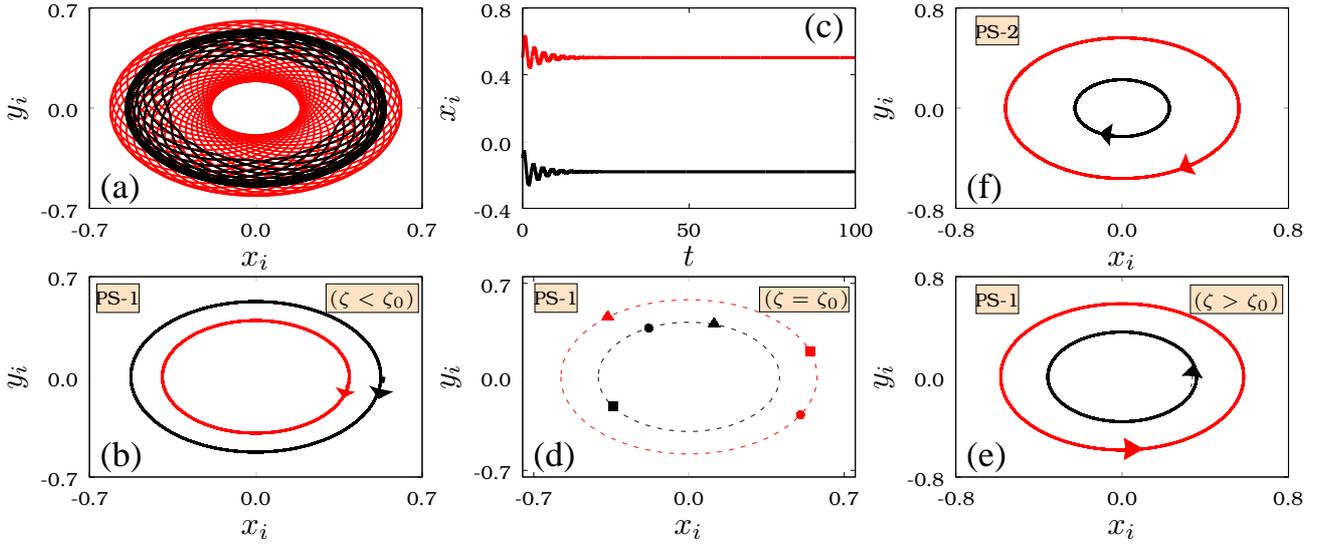}
		\caption{Dynamics of the system for different $\zeta$ values and for $\gamma_1=0.25$, $\gamma_2=1.0$ and $\omega_1=\omega_2=1.0$ are shown in the above figures. Figs. \ref{fig2}(a) and \ref{fig2}(b) are plotted respectively for two different initial conditions with $\zeta=0.56$ and they provide the phase portraits of the first and second oscillator. Figs. \ref{fig2}(c) and \ref{fig2}(d) are plotted for $\zeta=\zeta_0=1.22899$.  In Fig. \ref{fig2}(d), we have considered three different initial conditions and have given the final asymptotic state corresponding to these initial conditions by filled circle, square and triangle.  The dashed circles in Fig. 2(d) represent the circles in which the steady states or fixed points of the system are distributed at $\zeta=\zeta_0$. Figs. \ref{fig2}(e) and \ref{fig2}(f) are plotted at $\zeta=2.1$ to illustrate the multistability between clockwise and anti-clockwise oscillations.  These figures are plotted for two different initial conditions through which stabilization of the system toward PS$-1$ and PS$-2$ states is illustrated.  In all these figures, red curve (or red coloured attributes) represents the states of the first oscillator and black curve (or attributes in black colour) represents the state of the second oscillator.  The arrows in the phase potraits (Figs. \ref{fig2}(b), \ref{fig2}(e) and \ref{fig2}(f)) represent the direction of oscillation.}
		\label{fig2}
	\end{center}
\end{figure*}

 From Figs. \ref{fig1}(a)-\ref{fig1}(d), we can see that there are four different dynamical regions (I-IV).  In region-I, there are no stable periodic states as seen in Fig. \ref{fig1}(a) or \ref{fig1}(b) and the numerical studies show the existence of stable quasi-periodic or torus type oscillations in this region.  The stabilization of periodic states are found only at $\zeta \equiv 0.54$, that is in region-II where the green curve 
(colour online) with filled triangles represent a stable periodic state.  We have labeled the latter periodic states as PS$-1$.  We can observe from Fig. \ref{fig1}(a) and Fig. \ref{fig1}(b) that even in the region-II, the torus oscillations are found to be stable and so there exists multi-stability between PS$-1$ state and quasi-periodic states in the region-II.  The phase portraits of the coexisting torus oscillation and PS$-1$ periodic orbit at $\zeta=0.56$ are shown respectively in Figs. \ref{fig2}(a) and \ref{fig2}(b).   For $\zeta>0.59$, the torus oscillations are no more stable and PS$-1$ state alone is stable in region-III as observed in Figs. \ref{fig1}(a) and \ref{fig1}(b).  The latter figures also show the stabilization of one more periodic state, namely PS$-2$, in region-IV (that is for $\zeta>2.05$) and so there will be a multi-stability in the region-IV. Both PS-1 and PS-2 states are found to be stable for all higher values of $\zeta>2.05$.
\par Being discussed the stabilization of different periodic and quasi-periodic states, let us now have a closer look at their dynamical characteristics with respect to $\zeta$.   It is clear from Eq. (\ref{rat1}) that in the presence of coupling, all the periodic states including stable and unstable ones mentioned in Figs. \ref{fig1}(a)-\ref{fig1}(d) correspond to second order synchronized state (where $\dot{\theta}_2^*=2 \dot{\theta}_1^*$).  However, in the isolated case $\zeta=0$, the two oscillators are found to oscillate with identical amplitude and frequency.  So, the system does not stabilize in any second order synchronized periodic state as soon as the coupling is introduced and quasi-periodic oscillations appear in the initial transition region.  For a slightly higher value of $\zeta$ ($\zeta=0.54$), the stabilization of a second order synchronized state, namely PS$-1$ occurs and this state is found to be stable for all values of $\zeta>0.54$.  A closer look at Figs. \ref{fig1}(a) and \ref{fig1}(b) tells us that the amplitude of the second oscillator is higher than the amplitude of the first oscillator near $\zeta=0.54$ and with the increase of $\zeta$, an opposite behaviour is observed where the amplitude of the first oscillator becomes higher.  From Eq. (\ref{rat1}), it is clear that the frequency of the second oscillator is (two times) higher than the first one.  We infer here that the amplitude of the second oscillator is quickly adjusted to have a smaller value than the first oscillator so that it can have higher frequency of oscillations than the first oscillator.  Now considering the frequency of oscillation corresponding to the state PS$-1$, Fig. \ref{fig1}(c) shows an interesting phenomenon where the angular frequency $\dot{\theta}_1^*$ (or equivalently $\displaystyle{\frac{\dot{\theta}_2^*}{2}}$) changes its sign at the point $\zeta_0$.  
\par  For $\zeta< \zeta_0$, $\dot{\theta}_1^*$ (also $\dot{\theta}_2^*$) is found to be negative and so we observe oscillations in the clockwise direction as shown in Fig. \ref{fig2}(b) (we recall that even in the uncoupled situation the angular frequencies $\dot{\theta}_1^*$ and $\dot{\theta}_2^*$ ($=-\omega$) of the oscillators are negative which represent clockwise evolution).  At the point $\zeta=\zeta_0$, $\dot{\theta_1}^*$ and $\dot{\theta_2}^*$ becomes zero and the value of $\zeta_0$ is given by
{\small\begin{eqnarray}
\zeta_0=\left[\left(\frac{1}{4k}\right) \times \left(\gamma_1 \gamma_2 (2 \omega_2-\omega_1)^2+\frac{\gamma_2}{\gamma_1} \omega_1 (\omega_1^2+2 \omega_2^2)^2\right) \right]^{1/2}, \nonumber\\
\label{z}
\end{eqnarray}}
where $k=(\omega_1+\omega_2)(\omega_1^2+2 \omega_2^2)$.
The dynamics at the considered parametric values $\gamma_1=0.25$, $\gamma_2=1.0$ with $\omega_1=\omega_2=1.0$ and $\zeta_0=1.22899$ is shown in Fig. \ref{fig2}(c).  Due to the zero frequency nature, non-oscillatory or steady state dynamics is observed in the latter figure.  Secondly, the stabilization does not occur at a particular point. Instead of that the system stabilizes towards one of the fixed points distributed over a circular orbit.  In other words, for a particular initial condition, the first oscillator will stabilize to any one of the points that lie over a circular orbit of radius
\begin{eqnarray}
R_1^*=\sqrt{\frac{\gamma_1}{\gamma_2} \frac{2 \omega_2(\omega_1+\omega_2)}{(\omega_1^2+2 \omega_2^2)}} \label{er1}
\end{eqnarray} 
and the second oscillator will stabilize towards a point in the circular orbit of radius
\begin{eqnarray}
R_2^*=\sqrt{\frac{\gamma_1}{\gamma_2} \frac{\omega_1(\omega_1+\omega_2)}{(\omega_1^2+2 \omega_2^2)}}, \label{er2}
\end{eqnarray}
where the positions of the two oscillators in these circular orbits are restricted to have the phase difference
\begin{eqnarray}
\delta^*=\theta_2^*-2 \theta_1^*=-\frac{\omega_1}{2 \zeta R_2^*}. \label{epd}
\end{eqnarray} 
To illustrate the above, we consider a different initial condition say $\zeta_0=1.22899$. In this case, the stabilization of the system towards different points along the circular orbits of radius $R_1^*$ and $R_2^*$ is illustrated in Fig. \ref{fig2}(d).  In this figure, we have considered three different initial conditions and represent the final asymptotic states of the two oscillators by the filled circle, square and triangle.  
\par Upon increasing the value of $\zeta$ beyond $\zeta_0$, one may observe from Fig. \ref{fig1}(c) that the frequency starts increasing and once $\dot{\theta}_1^*$ (and $\dot{\theta}_2^*$) become positive then the system execute oscillations in the anticlockwise direction.  For instance, for $\zeta=2.1$, the existence of anticlockwise oscillation is shown in Fig. \ref{fig2}(e).   Thus, on the whole, the Figs. \ref{fig2}(b), \ref{fig2}(d) and \ref{fig2}(e) clearly illustrate the change of clockwise rotation in the PS$-1$ oscillatory state to anti-clockwise direction upon increasing the value of $\zeta$.  From these three figures, we can also observe that initially the orbit of the second oscillator is higher than the first oscillator - see Fig. \ref{fig2}(b) (which is plotted for $\zeta=0.56$). Now upon increasing the value of $\zeta$, the periodic orbit of the first oscillator becomes larger than the second one which can be noticed from  Figs. \ref{fig2}(d) and \ref{fig2}(e) as we have mentioned in subsection \ref{III_a}.
\begin{figure*}[htb!]
	\centering
	\includegraphics[width=0.85\linewidth]{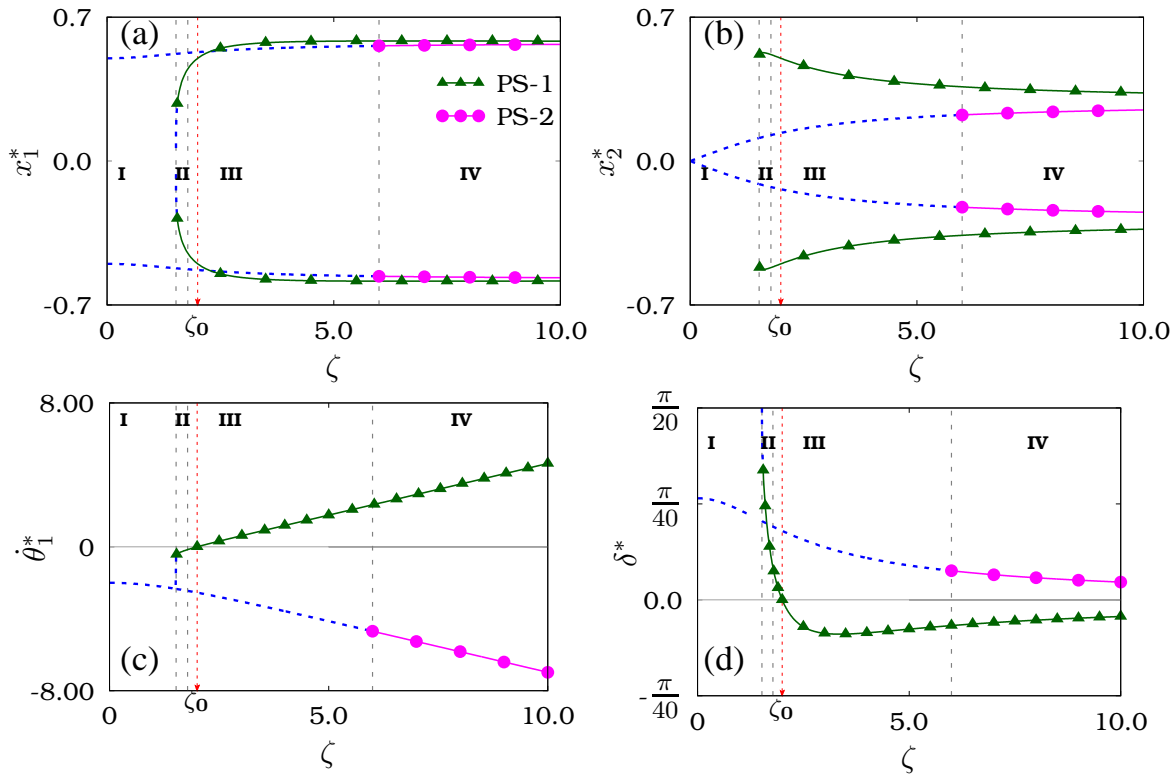}
	\caption{Maxima and minima values of the variables $x_1$ and $x_2$ in different stable and unstable stationary states of the system for $\omega_1 \neq \omega_2$ are shown respectively in Figs. \ref{fig3}(a) and \ref{fig3}(b) for different values of $\zeta$.  Similarly, the angular frequency $\dot{\theta_1}^*$ and the relative phase difference $\delta^*$ corresponding to different stable and unstable states are shown with respect to $\zeta$ in Figs. \ref{fig3}(c) and \ref{fig3}(d).  In all the above figures, we represent the unstable states with dashed lines and the stable states with continuous line with some filled  attributes like filled circle or filled triangles. In all these figures, we consider $\gamma_1=0.25$, $\gamma_2=1.0$ and {\bf$\omega_1=2$, $\omega_2=1$}. }
	\label{fig3}
\end{figure*}
\par As far as PS$-2$ state is concerned, Figs. \ref{fig1}(a) and \ref{fig1}(b) tell us that throughout its stable regime, the amplitude of the second oscillator is smaller than the first oscillator and Fig. \ref{fig1}(c) indicates that $\dot{\theta}_1^*$ (also $\dot{\theta}_2^*$) remain negative throughout its stable regime so that the two oscillators evolve in clockwise direction.  As we noted earlier, in the stable region of PS$-2$ state (that is, in the region-IV), PS$-1$ state is stable but here PS$-1$ state executes oscillations in the anti-clockwise direction and PS$-2$ state executes oscillations in the clockwise direction.  Hence we can observe a multistability in region-IV due to the coexistence of clockwise and anticlockwise oscillations.  This has also been illustrated in the Figs. \ref{fig2}(e) and \ref{fig2}(f) where the initial conditions leading to PS$-1$ show anticlockwise rotation (Fig. \ref{fig2}(e)) and the ones leading to PS$-2$ exhibit clockwise rotation (Fig. \ref{fig2}(f)). 
\par In summary, the present case brought out certain important features of the introduced asymmetric coupling where it not only facilitates second order synchronization but also induce changes in the direction of oscillations thereby it gives rise to multi-stability of clockwise and anticlockwise oscillatory states. 
\subsection{Case (ii): $\omega_1\ne\omega_2$}{\label{III_b}}
In the previous sub-section, we discussed the role of asymmetric coupling in the case of identical oscillators.  In this sub-section, we consider coupling among non-identical oscillators and bring out their dynamical characteristics.   In this connection, we plot Fig. \ref{fig3} where the stable and unstable nature of the periodic orbits are presented along with the details of their amplitude, phase and frequency.   The Fig. \ref{fig3} makes it clear that the dynamical characteristics of the present case is similar to that of the previous case.  For instance, Figs. \ref{fig3}(a)-\ref{fig3}(d) show the existence of torus oscillations for smaller coupling strengths, that is before the stabilization of second order synchronized state.  A periodic state, namely PS$-1$ is stabilized at the boundary of region II and in the latter region, multistability between torus and PS$-1$ states do exist. In region III, torus oscillations lose their stability and Fig. \ref{fig3}(c) shows that at the point $\zeta_0=2.0$ (the value of $\zeta_0$ matches with the expression given in Eq. (\ref{z})), there occurs a transition from clockwise to anti-clockwise oscillation.  In region IV, as similar to the previous case, stabilization of a new periodic state, namely PS$-2$ occurs which gives rise to multistability of clockwise and anti-clockwise oscillations in this region. 

\par
\begin{figure*}[h]
  \begin{center}
			\includegraphics[width=0.95\linewidth]{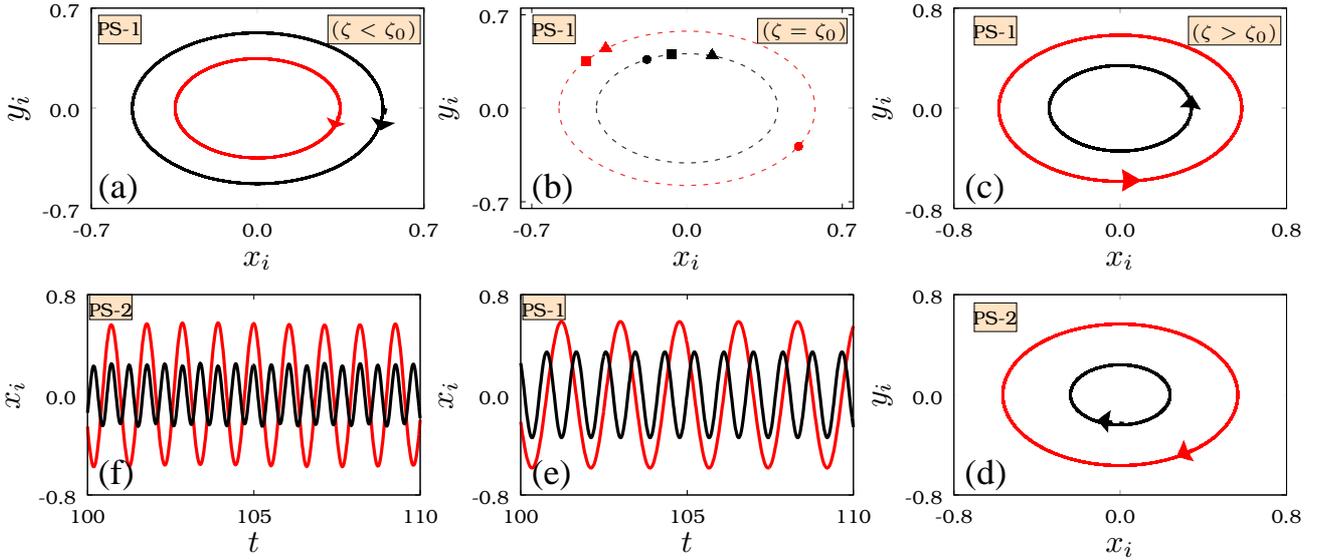}
			\caption{Dynamics of the system for different $\zeta$ values and for $\gamma_1=0.25$, $\gamma_2=1.0$ and $\omega_1=2, \omega_2=1.0$. Figs. \ref{fig_dyn2}(a)-\ref{fig_dyn2}(d) are the phase portraits for first and second oscillators for different values of $\zeta$ with (a) $\zeta=1.6<\zeta_0$ (in region-2), (b) $\zeta=\zeta_0=2.0$ (region-3) and (c) and (d) $\zeta=8.1>\zeta_0$ (region-4) respectively. Figs. \ref{fig_dyn2}(a)-\ref{fig_dyn2}(c) illustrate the change in direction of oscillation in PS-1 state.  To illustrate steady state dynamics at critical value, we considered three different initial conditions in Fig. \ref{fig_dyn2}(b) and have given final asymptotic states respectively by circle, square and triangle.  In this, the final asymptotic states of the first and second oscillators lie respectively over the red and black (dashed) circles. Figs. \ref{fig_dyn2}(c)-\ref{fig_dyn2}(f) illustrates the multistability of PS-1 and PS-2 states in region-4 ($\zeta=8.1$) where Figs. \ref{fig_dyn2}(c) and \ref{fig_dyn2}(e) (Figs. \ref{fig_dyn2}(d) and \ref{fig_dyn2}(f)) represent the phase portrait and temporal dynamics of PS-1 (PS-2) state in region-4.   In all these figures, red curve (or red coloured attributes) represents the states of the first oscillator and black curve (or attributes in black colour) represents the state of the second oscillator.  The arrows in the phase potraits (Figs. \ref{fig_dyn2}(a), \ref{fig_dyn2}(c) and \ref{fig_dyn2}(d)) represent the direction of oscillation.}
			\label{fig_dyn2}
  \end{center}
\end{figure*}

 Thus, it is more clear from Fig. \ref{fig3} that as similar to the previous case, we can observe the same dynamical features like stabilization of $1:2$ synchronized state for stronger coupling strengths and coexisting clockwise and anticlockwise oscillations.  We plot the dynamics of the system corresponding to this case in Fig. \ref{fig_dyn2}. Figures \ref{fig_dyn2}(a)-\ref{fig_dyn2}(c) represent the changes in the direction of oscillation in PS-1 state.  Figure \ref{fig_dyn2}(a) elucidates the stabilization of PS-1 state in the region-II and for the choice $\zeta=1.6<\zeta_0$, the oscillations are in the clockwise direction.  While varying $\zeta$, at $\zeta=\zeta_0$, due to the change in the direction of oscillation, the frequency of the oscillations become zero and the oscillators execute steady state dynamics as shown in Fig. \ref{fig_dyn2}(b). 
As seen in the previous case (Figs. \ref{fig2}(c) and \ref{fig2}(d)), the system has infinite number of fixed points that lie over the circles (dashed curve in Fig. \ref{fig2}(b)) which are defined by the radii ($R_1^*$ and $R_2^*$) and phase difference ($\delta^*$) as given in Eqs. (\ref{er1})-(\ref{epd}).  To elucidate the above, we consider three different initial conditions in Fig. \ref{fig_dyn2}(b) and the final asymptotic states are represented by filled circle, square and triangle.  For $\zeta=8.1>\zeta_0$ (which lies in the region-4), the multistability of PS-1 (anticlockwise) and PS-2 (clockwise) oscillations are shown in Figs. \ref{fig_dyn2}(c) and \ref{fig_dyn2}(d) respectively.  The corresponding temporal dynamics in PS-1 and PS-2 states are given in Figs. \ref{fig_dyn2}(e) and \ref{fig_dyn2}(f) and these two figures make clear that in both PS-1 and PS-2 states, the frequencies of the two oscillators are not same and from Eq. (\ref{rat1}) we can confirm that these periodic states (PS-$1$ and PS-$2$) are in $1:2$ synchronized state.  In the following, we consider the case where the natural frequencies of the uncoupled oscillators have the frequency relation $\omega_1:\omega_2=1:2$ and investigate the role of asymmetric coupling in such a case.
\section{\label{sec4}Case \lowercase{(iii)}: $\omega_2=2\omega_1$}
As we did earlier, in this interesting limiting case also, we first analytically study the dynamics of the system.  For this purpose, we consider $\alpha=R_1\exp{(i\theta_1)}$ and $\beta=R_2\exp{(i\theta_2)}$, where we take $\theta_1=-\omega_1t+\phi_1$ and $\theta_2=-\omega_2 t+\phi_2$.  The resultant amplitude and phase equations are turned out to be
\begin{eqnarray}
\dot R_1&=&(\gamma_1-\gamma_2R_1^2)R_1+2\zeta R_1R_2\sin (\phi_2-2 \phi_1), \label{rk1}\nonumber\\
\dot R_2&=&(\gamma_1-\gamma_2R_2^2)R_2-\zeta R_1^2\sin (\phi_2-2 \phi_1), \label{rk2}\nonumber\\
\dot{\phi}_1&=&-2\zeta R_2\cos(\phi_2-2\phi_1),\label{phik1} \nonumber \\
\dot{\phi}_2&=&-\zeta \frac{R_1^2}{R_2}\cos(\phi_2-2\phi_1). \label{phik2}
\end{eqnarray}
\begin{figure*}[htb!]
	\centering
	\includegraphics[width=0.85\linewidth]{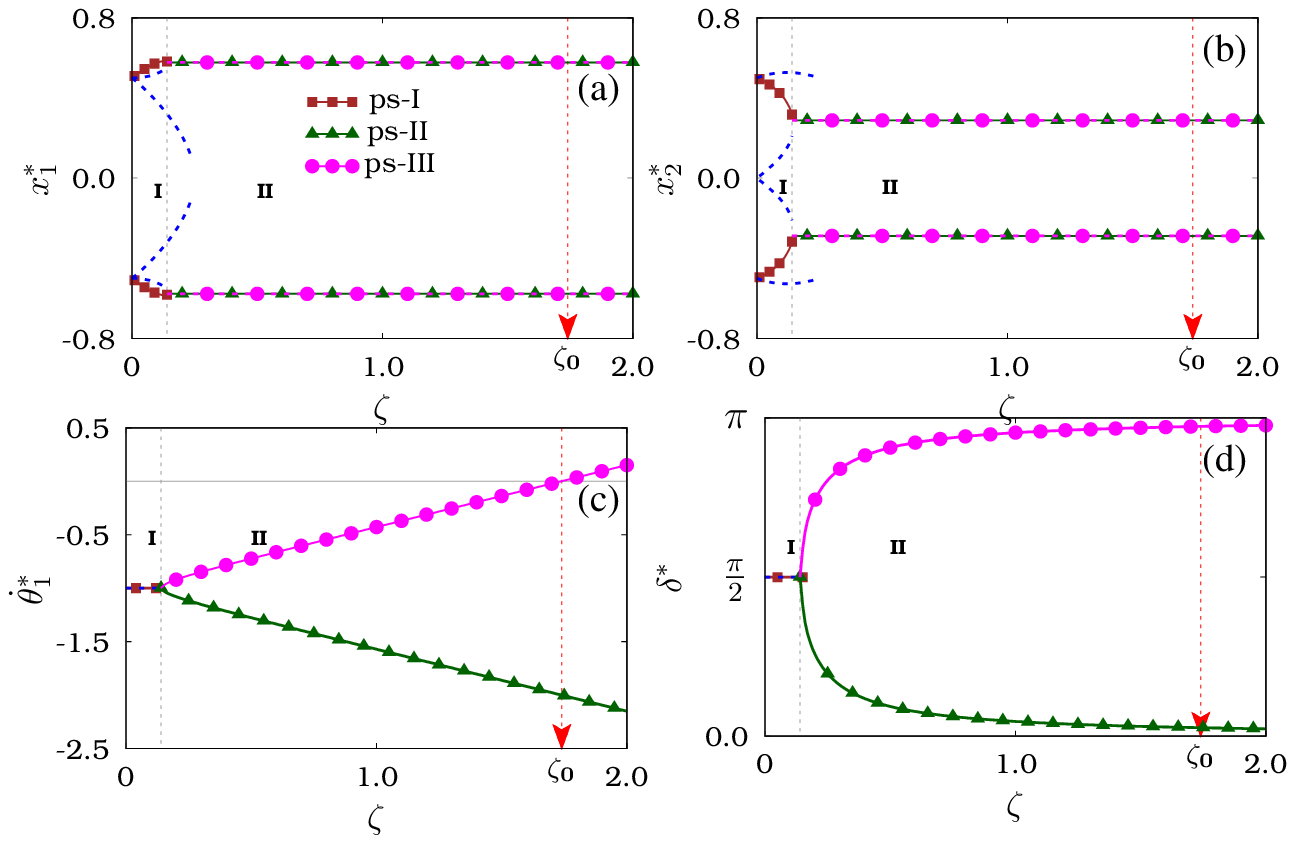}
	\caption{Maxima and minima values of the variables $x_1$ and $x_2$ in different stable and unstable stationary states of the system for $\omega_2=2\omega_1$ are shown respectively in Figs. \ref{fig4}(a) and \ref{fig4}(b) for different values of $\zeta$.  Similarly, the angular frequency $\dot{\theta_1}^*$ and the relative phase difference $\delta^*$ corresponding to different stable and unstable states are shown with respect to $\zeta$ in Figs. \ref{fig4}(c) and \ref{fig4}(d).  In all the above figures, we represent the unstable states with dashed lines and the stable states with continuous line with some filled  attributes like filled circle or filled triangles. In all these figures, we consider $\gamma_1=0.25$, $\gamma_2=1.0$ and {\bf$\omega_1=1$, $\omega_2=2$}. }
	\label{fig4}
\end{figure*}
\par We reduce the above equations to the form
\begin{eqnarray}
\dot R_1&=&(\gamma_1-\gamma_2R_1^2)R_1+2\zeta R_1R_2\sin\delta, \label{r1}\nonumber\\
\dot R_2&=&(\gamma_1-\gamma_2R_2^2)R_2-\zeta R_1^2\sin\delta, \label{r2}\nonumber\\
\dot\delta&=&-\zeta\big(\frac{R_1^2-4R_2^2}{R_2}\big)\cos\delta,\label{delta3}
\end{eqnarray}
with $\delta=\theta_2-2\theta_1=\phi_2-2\phi_1$.  We now trace out the possible stationary states or periodic orbits corresponding to this case. From the stability analysis we found that there exists two regions  where the system exhibits different dynamical behaviours. The dynamics of the stable and unstable periodic orbits in these dynamical regions are given below. 
\begin{enumerate}[(i)]
	\item For lower values of $\zeta$ we find that periodic orbits corresponding to $\delta=\pi/2$ and $\delta=3\pi/2$ do exist.
	When $\delta=\pi/2$, the amplitude of the corresponding state takes the following value 
	\begin{equation}
	R_1^*=\sqrt{\frac{\gamma_1}{\gamma_2}+\frac{2\zeta R_2^*}{\gamma_2}},
	\end{equation} 
	in which $R_2^*$ is real and takes positive valued solutions (as $R_1$, $R_2$ $\in$ ${\cal{R}}^+$) from the following polynomial 
	\begin{equation}
	R_2^{*3}-\frac{\gamma_1}{\gamma_2}R_2^*+\frac{\zeta}{\gamma_2} R_1^{*2}=0. \label{R2po_3}
	\end{equation}
	\item Similarly, when $\delta=3 \pi/2$
	\begin{equation}
	R_1^*=\sqrt{\frac{\gamma_1}{\gamma_2}-\frac{2\zeta R_2^*}{\gamma_2}},
	\end{equation} 
	and $R_2^*$ is the solution of the polynomial
	\begin{equation}
	R_2^{*3}-\frac{\gamma_1}{\gamma_2}R_2^*-\frac{\zeta}{\gamma_2} R_1^{*2}=0. \label{R2po_4}
	\end{equation}
	As $\cos \delta=0$ for these states with $\delta=$ $\pi/2$ and $3 \pi/2$, $\dot{\phi}_1^*=\dot{\phi}_2^*=0$ and so 
	\begin{eqnarray}
	\dot{\theta_1}^*&=&-\omega_1+\dot{\phi_1}^*=-\omega_1 \nonumber \\   \dot{\theta_2}^*&=&-\omega_2+\dot{\phi_2}^*=-\omega_2=-2 \omega_1
	\label{dtk}
	\end{eqnarray} 
	Equation (\ref{dtk}) shows that the frequency of the two oscillators in the above states is same as their intrinsic frequencies and it does not vary with respect to $\zeta$.  Due to this reason, these stationary states represent second order synchronized state where $\dot{\theta}_1^*:\dot{\theta}_2^*=1:2$.
	\item For the higher coupling strength, $\zeta>\zeta_c=\sqrt{\gamma_1\gamma_2/12}$, a set of stationary periodic states with amplitude ratio $R_1^*:R_2^*=2:1$ do exist.  These two states have the same amplitude with 
	\begin{equation}
	R_1^*=2\sqrt{\frac{\gamma_1}{3\gamma_2}} \quad \mathrm{and} \quad R_2^*=\sqrt{\frac{\gamma_1}{3\gamma_2}},
	\label{sk}
	\end{equation}
	and they differ only by the phase difference. The phase difference  $\delta^*$ corresponding to these two states can be respectively given by
	\begin{eqnarray}
	\delta^*=\begin{cases}
	~~\delta_0 \\ -\delta_0+\pi, 
	\end{cases} \nonumber\\
	\text{with}~~ \delta_0=\sin^{-1}\left[\frac{1}{2\zeta}\sqrt{\frac{\gamma_1\gamma_2}{3}}\right].
	\end{eqnarray} \label{st_ph}
	Considering the frequencies of the two oscillators in these states, we find that in the state with $\delta^*=\delta_0$,
	\begin{eqnarray}
	\dot{\theta_1}^*&=&-\omega_1-2 \zeta \sqrt{\frac{\gamma_1}{3 \gamma_2}} \cos \delta_0,  \nonumber \\ \dot{\theta_2}^*&=&-2 \omega_1-4 \zeta \sqrt{\frac{\gamma_1}{3 \gamma_2}} \cos \delta_0,
	\label{frk1}
	\end{eqnarray}
	and in the state with $\delta^*=-\delta_0+\pi$
	\begin{eqnarray}
	\dot{\theta}_1^*&=&-\omega_1+2 \zeta \sqrt{\frac{\gamma_1}{3 \gamma_2}} \cos \delta_0, \nonumber \\   
	\dot{\theta}_2^*&=&-2 \omega_1+4 \zeta \sqrt{\frac{\gamma_1}{3 \gamma_2}} \cos \delta_0.
	\label{frk2}
	\end{eqnarray}
\end{enumerate}
\par Thus the two states mentioned by $\delta^*=\delta_0$ and $-\delta_0+\pi$ do posses the same amplitudes but have different frequencies as given in Eqs. (\ref{frk1}) and (\ref{frk2}).   Importantly, the latter two equations indicate that these two states also have the frequency ratio in the order $\dot{\theta}_1^*:\dot{\theta}_2^*=1:2$ and so they also represent the second order synchronized state.  Yet another important observation that we can have from Eqs. (\ref{frk1}) and (\ref{frk2}) is that in the case $\delta^*=\delta_0$, $\dot{\theta}_{1,2}^*$ remain negative (representing clockwise oscillations) for $\zeta<\zeta_0$, where $\zeta_0$ can be obtained from the expression (\ref{z})
and they take positive values (representing anti-clockwise oscillations) when $\zeta>\zeta_0$.  But in the case $\delta^*=-\delta_0+\pi$, Eq. (\ref{frk2}) shows that $\dot{\theta}_{1,2}^*$ remain negative for all values of $\zeta$ and so they represent clockwise oscillations.  
\par In the above, we discussed different stationary periodic orbits of the system. Now we  turn our attention to the stability nature of the periodic states.  In this connection, we plot Fig. \ref{fig4} by analyzing the existence and stability nature of the above stationary periodic orbits for $\omega_1=1.0$ and $\omega_2=2.0$.   As mentioned earlier there are two different dynamical regions (I-II) as illustrated in Figs. {\ref{fig4}}(a)-(d) and for smaller values of $\zeta$ (that is, for $\zeta<\zeta_c$), the states corresponding to $\delta^*=\pi/2$ and $3 \pi/2$ (given in Eqs. (\ref{R2po_3}) and (\ref{R2po_4})) only do exist.  For the considered case ($\omega_1=1.0$ and $\omega_2=2.0$), the polynomial in Eq. (\ref{R2po_3}) corresponding to $\delta^*=\pi/2$ yields two positive real roots while Eq. (\ref{R2po_4}) corresponding to $\delta^*=3 \pi/2$ yields only one root. Thus, three stationary periodic orbits are possible for the lower values of $\zeta$.  Among the three, only one of the states corresponding to $\delta^*=\pi/2$ is found to be stable which is represented by the brown curve (colour online) with filled squares and the other two unstable periodic states are represented by dashed blue curves (colour online) in Figs. {\ref{fig4}}(a)-(d).   Increasing the $\zeta$ value beyond $\zeta_c$ (that is, in the region-II), the stable branch of $\delta^*=\pi/2$ becomes unstable and the states corresponding to $\delta^*=\delta_0$ and $-\delta_0+\pi$ comes to existence.   The stability of the latter two states are studied and their stable region is obtained with the use of Routh-Hurwitz criterion \cite{ml}.  The result shows that these two states become stable as soon as they come into existence (that is, for $\zeta>\zeta_c=\sqrt{\gamma_1\gamma_2/12}=0.14$).  The stable branches of these two states are respectively indicated by green curve (color online) with filled triangles (ps-II state) and magenta curve (color online) with filled circles (ps-III state) in Figs. \ref{fig4}(a)-\ref{fig4}(d).
\par
\begin{figure*}[htb!]
	\begin{center}
		\includegraphics[width=1.01\linewidth]{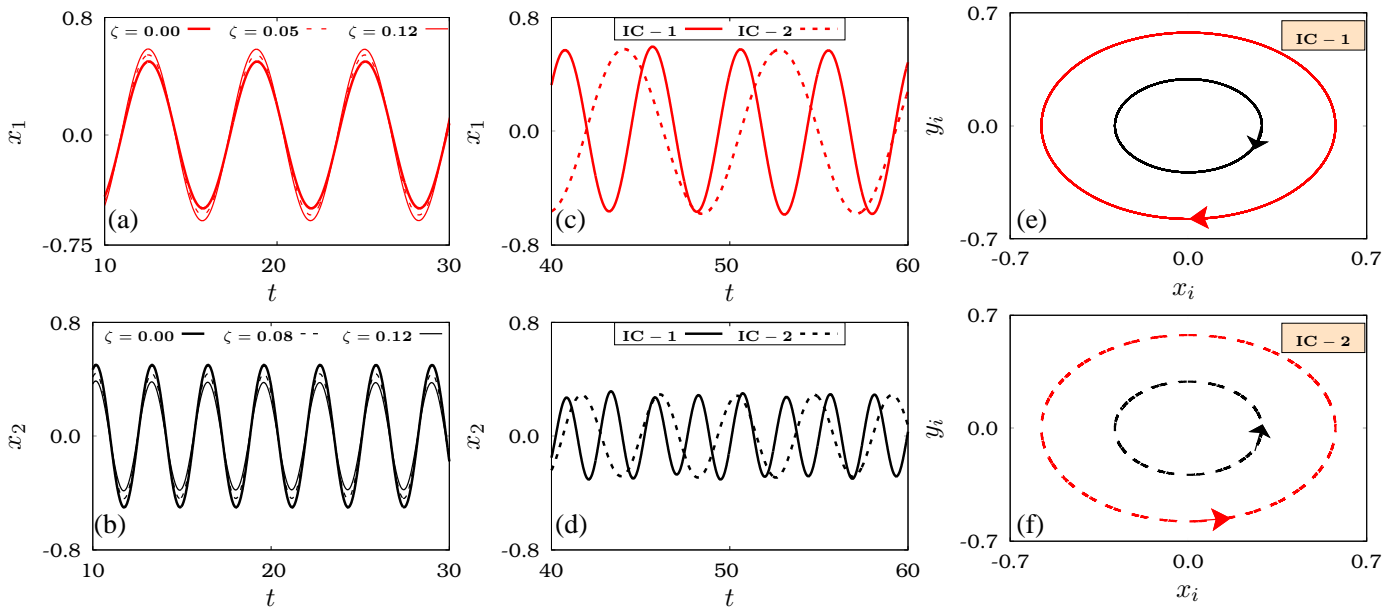}
		\caption{Dynamics of the system for different $\zeta$ values and for $\gamma_1=0.25$, $\gamma_2=1.0$ and $\omega_1=1$ and $\omega_2=2.0$ are shown in the above figures. Figs. \ref{fig5}(a) and \ref{fig5}(b) are plotted for different values $\zeta$. Figs. \ref{fig5}(c) and \ref{fig5}(d) are plotted for $\zeta=\zeta_0=1.8$ for two different initial conditions.  Figs. \ref{fig5}(e) and (f) are plotted at $\zeta=1.8$ for two different initial conditions where the stabilization of the system toward ps-II and ps-III states is illustrated.  In all these figures, red curve (or red colored attributes) represents the states of the first oscillator and black curve (or attributes in black color) represents the state of the second oscillator.  The arrows in the phase portraits (Figs. \ref{fig5}(e) and \ref{fig5}(f)) represent the direction of oscillation.}
		\label{fig5}
	\end{center}
\end{figure*}

 With this knowledge on the stable states of the system for different values of $\zeta$, in the following, we proceed to discuss the dynamical characteristics of the system.  The first difference which we notice in this case (from Fig. \ref{fig4}) in comparison to the previous cases is that the missing of torus oscillation.  As the intrinsic frequency itself has the relation $\omega_1:\omega_2=1:2$, there is no need for torus oscillation to mediate the stabilization of second order synchronized periodic state.  Thus the periodic state ps-I is stable for lower coupling strengths ($\zeta<\zeta_c$, region-I).  By looking at the branch corresponding to this ps-I state in Figs. \ref{fig4}(a) and \ref{fig4}(b), we understand that while we increase the value of $\zeta$ from zero (in region-I), the amplitude of the first oscillator increases while the counterpart of the second oscillator decreases. We note that in the absence of coupling, the two oscillators oscillate with the same amplitude. Correspondingly, Figs. \ref{fig4}(c) and \ref{fig4}(d) indicate that for all values of $\zeta<\zeta_c$ (in region-I), the frequency of the two oscillators remains the same as that of the natural frequency (also shown by Eq. (\ref{dtk})) and the $\delta$ value also remains the same as $\pi/2$.  The dynamics of the two oscillators in this region-I are presented respectively in Figs. \ref{fig5}(a) and \ref{fig5}(b) for $\zeta=0.0,$ $0.05$ and $0.12$.  From these figures, we observe that the amplitude of oscillations alone changes with respect to $\zeta$ and frequencies remain the same as that of the $\zeta=0.0$ case.    
\par  As mentioned earlier, while $\zeta<\zeta_c$, the amplitude of the two oscillators varies with respect to $\zeta$ and it reaches the value  $R_1^*=2 R_2^*=2\sqrt{\displaystyle{\frac{\gamma_1}{3\gamma_2}}}$  at $\zeta=\zeta_c$. At this critical point ($\zeta=\zeta_c$), Fig. \ref{fig4} indicates the occurence of pitchfork like bifurcation where the two new states, namely ps-II and ps-III gain stability and ps-I loses its stable nature.  Beyond this critical point, the amplitudes of both the oscillators do not vary and it remains the same as $R_1^*=2 R_2^*=2\sqrt{\displaystyle{\frac{\gamma_1}{3\gamma_2}}}$ (vide Eq. (\ref{sk})) which can be seen from the stable branches of ps-II and ps-III in Figs. \ref{fig4}(a) and the \ref{fig4}(b).  The amplitude of oscillations in both the states ps-II and ps-III are found to be same and so the branches of ps-II and ps-III lie over each other in Figs. \ref{fig4}(a) and \ref{fig4}(b).  But considering the frequencies in ps-II and ps-III states, Eqs. (\ref{frk1}) and (\ref{frk2}) and Fig. \ref{fig4}(c) indicate that they are different and they vary with respect to $\zeta$.  Secondly, Eqs. (\ref{frk1}) and (\ref{frk2}) indicate that both the states represent second order synchronized state. Thus for all values of $\zeta$, frequency of the two oscillators (in both ps-II and ps-III states) are in the ratio $1:2$.   Comparing the amplitude and frequency ratios of these states ($R_1^*:R_2^*=2:1$, $\dot{\theta}_1^*:\dot{\theta}_2^*=1:2$ ) we find that the periodic orbit of the first oscillator is two times larger than the one corresponding to the second oscillator and the frequency of the second oscillator is two times than that of the first oscillator.  
\begin{figure}[htb!]
\begin{center}
		\includegraphics[width=0.8\columnwidth]{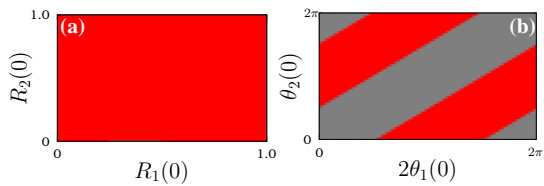}
		\caption{ On Figs.\ref{fig6}(a) and \ref{fig6}(b) we plot basin of attraction of phase and amplitude. In Fig. \ref{fig6}(a) we fix $\theta_1(0)=0.13$ and $\theta_2(0)=0.6$ and find the basin of attraction by varying $R_1(0)$ and $R_2(0)$. In Fig. \ref{fig6}(b) we fix $R_1(0)=0.1$ and $R_2(0)=2.0$ and determine basin of attraction by varying $\theta_1(0)$ and $\theta_2(0)$.}
		\label{fig6}
	\end{center}
\end{figure}
\par From Fig. \ref{fig4}(d), it is clear that the value of $\delta$ corresponding to these two states are different and they vary with respect to $\zeta$.  It is important to note that $\delta$ value of the ps-II state changes from positive to negative at the point $\zeta_0$ (in the considered case $\zeta_0=1.78$).  So, for $\zeta_c<\zeta<\zeta_0$, we have multistability between two clockwise oscillating periodic states and for $\zeta>\zeta_0$, we have multistability of clockwise and anticlockwise oscillations.  Another important thing which we come across in this multistable region is that the periodic orbits (or the amplitude) of the two multistable states are same while their corresponding frequencies and phase difference are different.  To illustrate it, we plot Figs. \ref{fig5}(c)-\ref{fig5}(f) for $\zeta=1.8>\zeta_0$.  In these figures, we show the dynamics of the system in ps-II  and ps-III  states by considering two different initial conditions.  The temporal behaviour presented in Figs. \ref{fig5}(c) and \ref{fig5}(d) illustrates that the oscillations corresponding to IC-1 and IC-2 are of same amplitude and they differ only by the frequency where the oscillation corresponding to IC-1 has frequency higher than that of IC-2.   The phase portraits corresponding to these two initial conditions are presented respectively in Figs. \ref{fig5}(e) and \ref{fig5}(f) which in turn confirms that the periodic orbit corresponding to ps-II and ps-III are same but ps-II represents clockwise oscillation ps-III represents anti-clockwise oscillations.  Thus, we can observe an interesting multistability in which  the system will show clockwise or anticlockwise oscillation (with different frequency) respectively for different initial conditions over the same periodic orbit.
\par To trace the initial conditions leading to ps-II and ps-III states in this multistable region, we plot the basin of attraction for $\zeta=0.3$.  For this purpose, we consider the initial conditions of the form $x_i(0)=R_i(0) \cos \theta_i(0)$ and $y_i(0)=R_i(0) \sin \theta_i(0)$, $i=1,2$. In Fig. \ref{fig6}(a), we fix $\theta_1(0)=0.13$, $\theta_2(0)=0.6$ and vary $R_1(0)$ and $R_2(0)$ to find the initial conditions leading to ps-II and ps-III states.  For the considered values of $\theta_i(0)$, we find the stabilization of ps-II state for all values of $R_1(0)$ and $R_2(0)$.  In Fig. \ref{fig6}(b), we fix $R_1(0)$ and $R_2(0)$ and vary $\theta_1(0)$ and $\theta_2(0)$.  The basin of attraction presented in Fig. \ref{fig6}(b) makes it clear that the stabilization of ps-II and ps-III states depend on the value of $\theta_i(0)$ and not on $R_i(0)$, $i=1,2$.
\section{\label{sec5}Conclusion}
\par In this work, we have demonstrated that due to the presence of a particular rotational symmetry in the coupled system, the nonlinear coupling enables high order ($1:2$) synchronization.   By considering different values for natural frequencies ($\omega_1$ and $\omega_2$), we have studied the dynamics in different cases. In the $\omega_2\neq2\omega_1$ case, we have shown that the system tries to adjust itself towards $1:2$ synchronized state and so torus type oscillations can be observed in weak coupling regime. In the strong coupling regime, we have observed that $1:2$ type synchronous periodic oscillations appear. We have also identified the multistability of clockwise and anticlockwise periodic states in this regime. In  the limiting case ($\omega_2=2\omega_1$) $1:2$ synchronous states are observed in all the parametric regions. In this case, we have demonstrated that the periodic orbit corresponding to clockwise and anticlockwise periodic states are the same and they differ only in phase and frequency. 
\par The higher order synchronization has been observed and applied in various biological and/or physical models \cite{schafer,velichi1,velichi2,velichi3,velichi4,velichi5,michaels,bychkov}. Hence we expect that the above mentioned possibility to induce high order synchornization may have interesting practical applications.

\section*{Acknowledgements}
NT wishes to thank National Board for Higher Mathematics, Government of India, for providing the Junior Research Fellowship under the Grant No. 02011/20/2018 NBHM (R.P)/R\&D II/15064. The work of MS forms part of a research project sponsored by Council of Scientific and Industrial Research, Government of India, under the Grant No. 03/1397/17/EMR-II.

\section*{Author contributions} All the authors contributed equally to the preparation of this manuscript.

\section*{Data Availability Statement} All datas generated or analyzed during this study are included in this article.

\end{document}